\documentclass[twocolumn]{bmcart}

\makeatletter
\renewenvironment{figure}[1][]{%
  \@float{figure}[#1]%
}{
  \end@float
}
\makeatother

\makeatletter
\renewenvironment{figure*}[1][]{%
  \@dblfloat{figure}[#1]%
}{
  \end@dblfloat
}
\makeatother

\usepackage{graphicx}

\usepackage{amsthm,amsmath}
\usepackage[utf8]{inputenc} 

\usepackage{siunitx} 
\usepackage[T1]{fontenc}
\usepackage[official]{eurosym}

\startlocaldefs
\endlocaldefs

\begin{document}

\begin{frontmatter}

\begin{fmbox}

\title{EEG classifier cross-task transfer to avoid training sessions in robot-assisted rehabilitation}

\author[
  addressref={aff1},                   
  email={niklas.kueper@dfki.de}   
]{\inits{N.K.}\fnm{Niklas} \snm{Kueper}}
\author[
  addressref={aff1},
  email={su-kyoung.kim@dfki.de}
]{\inits{S.K.}\fnm{Su Kyoung} \snm{Kim}}
\author[
  addressref={aff1, aff2},                   
corref={aff2},
  email={elsa.kirchner@uni-due.de; elsa.kirchner@dfki.de} 
]{\inits{E.K.}\fnm{Elsa Andrea} \snm{Kirchner}}\newline 


\address[id=aff1]{
  \orgdiv{Robotics Innovation Center},             
  \orgname{German Research Center for Artificial Intelligence (DFKI)},          
  \city{Bremen},                              
  \cny{Germany}                                    
} 
\address[id=aff2]{%
  \orgdiv{Institute of Medical Technology Systems},
  \orgname{University of Duisburg-Essen},
  \city{Duisburg},
  \cny{Germany}
}


\begin{abstract} 
\parttitle{Background} 
For an individualized support of patients during rehabilitation, learning of individual machine learning models from the human electroencephalogram (EEG) is required. 
Our approach allows labeled training data to be recorded without the need for a specific training session, which is important for the feasibility of using EEG to support exoskeleton-assisted therapy. For this, the planned exoskeleton-assisted rehabilitation enables bilateral mirror therapy, in which movement intentions can be inferred from the activity of the unaffected arm.
During this therapy, labeled EEG data can be collected to enable movement predictions of only the affected arm of a patient based on the EEG and a proposed transfer learning approach in the second step.

\parttitle{Methods} 
A study was conducted with 8 healthy subjects and the performance of the classifier transfer approach was evaluated. Each subject performed $3$ runs of $40$ self-intended unilateral and bilateral reaching movements toward a target while recording EEG data from $64$ channels, EMG data from $16$ channels, and motion tracking data. A support vector machine (SVM) classifier was trained under both movement conditions to make predictions for the same type of movement. Furthermore, the classifier was trained under the bilateral condition and then transferred to predict motion intention under the unilateral condition. 
The approach was evaluated comparing a custom EEG channel selection method to a standard electrode constellation under varying the number of EEG channels.  

\parttitle{Results} 
The results show that the performance of the classifier trained on selected EEG channels evoked by bilateral movement intentions is not significantly reduced compared to a classifier trained directly on EEG data including unilateral movement intentions. 
Comparisons with standard channel constellations show that the comparable performance was enabled by our knowledge-based channel selection. Moreover, the results show that our approach also works with only $8$ or even $4$ channels. 

\parttitle{Conclusion} 
It was shown that the proposed classifier transfer approach enables motion prediction without explicit collection of training data. Since the approach can be applied even with a small number of EEG channels, this speaks for the feasibility of the approach in real therapy sessions with patients and motivates further investigations with stroke patients.

\end{abstract}


\begin{keyword}
\kwd{EEG}
\kwd{movement prediction}
\kwd{rehabilitation}
\kwd{classifier transfer}
\kwd{robot assisted therapy}
\kwd{lateralized readiness potential (LRP)}
\kwd{event related potential (ERP)}
\kwd{BCI}
\kwd{transfer learning}
\end{keyword}


%
\end{fmbox}

\end{frontmatter}



\section*{Background}
With the demographic change, concepts for maintaining the health of an increasingly aging population must be rethought. As shown in a statistic by~\cite{luengo:2020economic} the cost of stroke in $2017$ was \euro{60} billion in the $32$ European countries alone. Necessary changes affect the entire range of health care from prevention and acute treatment to rehabilitation and reintegration of persons who are ill or injured into everyday life. One of the approaches pursued is the use of robots in rehabilitation therapy, which was clinically tested, for example, as early as $1994$ with the MIT-MANUS~\cite{krebs:1998robot} and meanwhile proven to improve arm and hand function and muscle strength~\cite{mehrholz:2018electromechanical} and to enable more effective therapy if combined with traditional physiotherapy \cite{fazekas:2019future}. The goals of using such complex technologies are manifold and range from reducing costs to increasing the efficiency of therapy and relieving the burden on the therapists, to achieve high repetitions in interactive and self-initiated therapy as well as to extend therapy options~\cite{poli:2013robotic}.  \\
On the other hand therapy strategies for the rehabilitation and restoration of functions in humans, as for example discussed in~\cite{huang:2020clinical} in order to maximally restore functions of patients with spinal cord injury (SCI), can be very intense to achieve optimal results. 
Regarding physiotherapy it was shown that patients receiving intensive peer mentoring during and after rehabilitation had greater gains in self-efficacy~\cite{gassaway:2017effects}, which is highly important in rehabilitation~\cite{kornhaber:2018resilience}. It further decreases the time for unplanned rehospitalizations~\cite{gassaway:2017effects}. \\

For support at home, robotic solutions provide assistance in daily living (ADL), which can range from helping a patient balance~\cite{kerdraon:2021evaluation} while walking and standing to complex assistance with reaching \cite{grimm:2016closed} and grasping~\cite{gerez:2020hybrid}. 
 
Active exoskeletons \cite{kirchner:2022towards} are commonly used for such assistance, as well as for rehabilitation therapies, and have shown to be effective in neuromotor rehabilitation, especially after stroke~\cite{noda:2012brain,hortal:2015using,singh:2021evidence}. %

The application of new robotic technologies, such as exoskeletons, gives us hope to cope up with higher demands caused by demographic changes, the orientation to more intensive therapy, self-efficacy of patients, relieve of therapists, extension of therapeutic options and reduction of cost, all together reducing the total disease burden by $6$ to $10$ per cent by $2040$~\cite{Mckinsey:2022ino}. \\
However, for an individualized support, learning of individual models from human data is required. To acquire such data without the demand of a person to attend long training and calibration sessions is very relevant. Here, we show that we can successfully combine rehabilitation sessions with data acquisition to train models that infer movement intentions in humans, which in turn, can then be supported by an assistive robotic device.

\subsection*{Exoskeletons and physiological measures in therapy}

To provide support when needed~\cite{pehlivan:2015minimal,mounis:2019assist,prange:2009systematic}, physiological measurements such as the electromyogram (EMG), which can for example be recorded from a healthy leg to control a disabled leg using an exoskeleton~\cite{zhang:2016development}, or the electroencephalogram (EEG) which allows conclusions to be drawn about a user's intention to move, are of great importance for successful neurorehabilitation~\cite{noda:2012brain,hortal:2015using}. In particular, EEG can be used to infer movement intentions~\cite{kirchner:2014multimodal,kirchner:2022towards}, for example, where a patient wants to move~\cite{kwak2015:lower,lee:2017brain}. There are many examples of how human EEG can be used to control exoskeletons using a brain computer interface (BCI)~\cite{Wolpaw:2002,kwak2015:lower,lee:2017brain,Soekadar:2015,noda:2012brain,hortal:2015using,folgheraiter:2012measuring}. However, BCIs are often not used to decode brain activity that correlates with the brain processes that control movement intention, planning, and execution; instead, other brain signals are used, such as the activity evoked in the visual cortex by flickering light, known as steady state visual evoked potentials (SSVEP)~\cite{kwak2015:lower,lee:2017brain,Lew:2012DetectionReaching} to rather artificially make use of the patient's EEG as a control input. To bridge the gap between brain and body caused by brain injury and to promote rehabilitation, such an approach is not the preferred one. Instead, brain activity that drives movement intention, planning, and execution should be used as a natural or intrinsic bridge between the brain and the body~\cite{Kirchner:2013towards,Kirchner:2013applicability}, using both, the physiological data that directly encodes the humans intention and the autonomous capabilities of the robotic system, i.e., an exoskeleton.

\subsection*{Transfer learning and classifier transfer in BCIs}

In the field of BCIs, long training sessions are often required to record a large amount of training data~\cite{Gemein:2020MLDiagnostics,Roy:2019DeepLearningEEG}.
However, transfer learning (TL) can be applied in BCI applications to reduce the calibration effort and training duration. This can be achieved by using prior knowledge or data that does not originate from the target session, subject or even measurement device or task~\cite{Wu:2020transfer}. TL has recently proven its potential to improve classification performance and reduce calibration times in several investigations, e.g.,~\cite{Zhang:2020application, Zou:2019inter, Wei:2016transfer, Li2020:transfer, Peterson2021:transfer}. This is of high importance for enhancing the usability of BCIs in real-world applications.

In addition to cross-session as well as cross-subject paradigms, a cross-task TL approach could achieve complete avoidance of calibration sessions and even enable learning in the first place if labeled data is available only for a training task but not for a test task, as in this work. 
However, in comparison, cross-task classification approaches based on EEG data and TL have been less comprehensively investigated~\cite{Wu:2020transfer}. Nevertheless, across different fields of BCI research, transfer approaches for cross-task EEG classification have been proposed. In our previous work, we showed that a classifier trained on EEG data from an observation scenario could be transferred to detect erroneous behavior of a robot during an interaction scenario~\cite{kim2013:classifier, kim2015:handling, kim2017:intrinsic,kim2020:flexible}. Therefore, the elicted event-related potential (ERP), namely the error-related potential (ErrP), could be classified in the transfer case although the tasks and ERP shapes differed between training and testing. The transferrability of a classifier for the detection of errors across tasks was also shown in~\cite{behncke2018:cross}, where a deep convolutional neural network was used to detect errors for two different error paradigms from intracranial EEG data. Another application of a classifier transfer is the detection of target and 
\textit{missed target }events from EEG while the classifier was trained on EEG data evoked by target and \emph{standard} events (oddball paradigm) to enhance the amount of available training data and, hence, to enable classification of EEG trials evoked during recognition of targets and the failure of recognition of those~\cite{woehrle2014:online, kirchner2019:transfer, kirchner2013:applicability}.
Further examples of applied cross-task EEG classification can be found in literature in the area of workload recognition~\cite{zhou2022:cross, kakkos2021:eeg, baldwin2012:adaptive}, where the performed workload task differed between training and testing a classifier or model. For example in~\cite{zhou2022:cross}, a domain adaptation approach was applied that improved the workload classification performance for the transfer case compared to a non-transfer case. Besides cross-task EEG classification, where a classifier is strictly trained on one task and tested on another task, fusion approaches were applied in which data from different tasks were combined for training a classifier~\cite{yang2023:cross,chen2017:enhancing}. 
However, classifier transfer and TL approaches for the detection of movement intentions in motor imagery or motor execution paradigms have been little investigated so far. A few examples are the decoding of forearm movements that were improved by combining data from motor imagery and motor execution tasks~\cite{lee2020:decoding} as well as the investigation of cross-subject and cross-task motor imagery classification~\cite{he2020:different}.  

Nevertheless, to our knowledge, the transferability of a classifier between bilateral and unilateral movements (upper body reaching movement tasks), especially conceptualised to support rehabilitation therapy, has not been proposed or investigated so far.

\subsection*{Approach and goal of the paper}  

In this work, we focus on how to generate training data for EEG-based intention recognition to guide support using an active exoskeleton~\cite{Kumar:2019ModularDesign} for unilateral arm movements of the affected arm after stroke. Our approach eliminates any time spent by the patient generating training data, as data collection is integrated into the therapy training itself. \\ 
For motion intention detection in human EEG, training data is needed to train a classifier or model for the prediction task. Even if a patient is able to complete such long training sessions, this is not desirable since all time available should be used for therapy in the early post-stroke period in which the brain is very plastic \cite{Fazekas:2019futureRoleRobots}. Waist of time in this very sensitive time by plainly recording data to train a classifier must be avoided as far as possible.\\
Here we present our approach to train a classifier during a therapy session that does not require intention recognition from EEG activity but makes use of the intelligence of the robotic system to timely map movement intention to movement execution. This is done by making use of the exoskeletons mirror mode in which a movement of the unaffected arm is transferred to the affected arm by the exoskeleton~\cite{Kumar:2019ModularDesign}. Hence the exoskeleton supports mirrored dual arm movements intended by the patient. While the patient is exercising in this mode, EEG data can be recorded. The recorded EEG contains activities associated with the planning and execution of bilateral arm movements and can be used to train a classifier to infer movement intention of both arms. The usability of such classifier to determine the motion intention of the affected arm alone, will be explained in the following methods section. To develop such a classifier-transfer approach and evaluate its feasibility in principle, we conducted a study with healthy subjects and report and discuss the results here. 

\section*{Methods}
\subsection*{Proposed Classifier Training Method}
\label{sec: Proposed Classifier Training Method}

\begin{table}[h!]
\renewcommand{\arraystretch}{1.3}
\setlength{\tabcolsep}{1.4em}
\centering
\caption{Definition of concept for training and testing the EEG classifier.}
\begin{tabular}{|c|c|c|c|}
\hline
\multicolumn{2}{|c|}{} & \textbf{training} & \textbf{testing}\\ \hline
\textbf{A} & \textbf{no transfer} & unilateral & unilateral\\ \hline
\textbf{B} & \textbf{no transfer} & bilateral  & bilateral\\ \hline
\textbf{C} & \textbf{transfer}    & bilateral  & unilateral\\ \hline
\end{tabular}
\label{tab:definition of train-test conditions}
\end{table}

To train an EEG-classifier to predict movement intentions only for the affected arm of stroke patients, the following two-step concept was developed. In the first step, the EEG classifier is trained during the execution of bilateral movements (in a mirror mode rehabilitation session). The onsets of the bilateral movements are infered from the non-affected arm to generate reliable labels to train the classifier. In the second step, the classifier is transferred to predict unilateral movements of the affected arm. The transfer approach consists of training on EEG data derived from bilateral movement executions and applying a custom EEG-channel selection to improve the transferability of the classifier by data adaptation. Since the LRP (Lateralized Readiness Potential)~\cite{DeJong:1988use,Gratton:1988pre}, that is associated with movement planning, can be observed from EEG-channels of the motor cortex side contralateral to the moved upper limb~\cite{Kutas:1980preparation}, we focus on the differences and similarities in the EEG data between bilateral and unilateral movement intentions. Therefore, we customly select EEG-channels for the data processing, that are related to the planning of unilateral movements with the affected limb. Hence, the abilities of the non-affected limb are used to generate reliable training labels and the classifier can be customly trained on the provided EEG-data, containing information about movement intentions of the affected arm. 
The proposed method was evaluated by conducting experiments, involving bilateral and unilateral movement tasks, executed by healthy subjects. The training and testing conditions of the evaluation are illustrated in Table \ref{tab:definition of train-test conditions}.

\subsection*{Experimental Setup and Procedure}

\begin{figure}
\includegraphics[width=82.5mm]{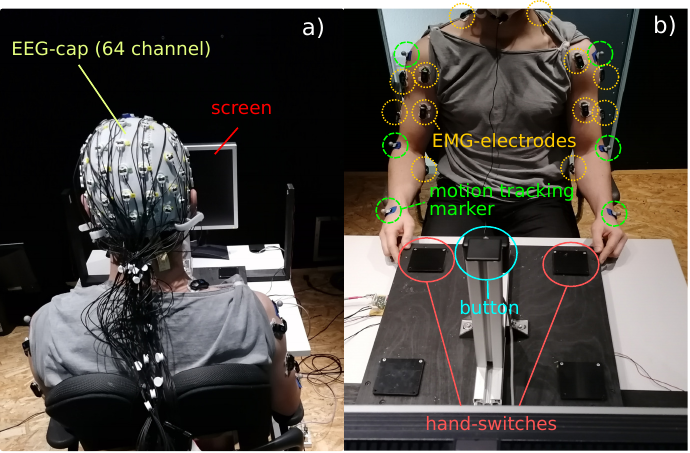}
\caption{Experimental setup of the study. In a) a subject is shown sitting in front of a screen wearing an EEG-cap with $64$ electrodes. In b) the custom build experimental board including hand-switches (orange) and a button (blue) as well as the placed EMG-sensors (yellow) and motion tracking marker (green) are illustrated.}
\label{Figure1}
\end{figure}

Eight healthy subjects ($4$ male, $4$ female) at the age of $25.5 \pm 4.0$ years participated in the conducted study. Only healthy right-handed subjects with no history of neurological or muscular diseases were recruited for the experiments. All subjects were advised to be well-rested for the experiment. The subjects were seated in a comfortable chair insight a shielded cabin. In front of the subjects a custom-build board, including hand-switches and a button was placed on a table. The subjects were asked to perform a reaching task, by pressing the button with their thumb. The button was placed at a height of approximately $25$ \si{\centi\meter} and at a distance of $30$ \si{\centi\meter} away from the resting position. The resting position was defined by the hand switches, where the subjects were asked to place their hands during the resting period. The position of the button was adjusted to the arm length of the subjects. Start and endpoint of the movements were standardized by ensuring a $90$-degree forearm-upper arm angle at rest and $0$ degree when pressing the button. \\ 
The experiment consisted of two different movement tasks which were $1)$ unilateral reaching movements and $2)$ bilateral reaching movements. The sequence of the task was varied between subjects (counterbalanced) in order to avoid possible learning effects. For the unilateral task, only the dominant right arm was moved whereby in the bilateral task a synchronous movement of both arms (both thumbs pressing the button) was executed. Each task included $3$ sets of $40$ self-initiated movements. Therefore, each subject performed a total of $120$ trials for each task. Each trial consisted of a resting period of at least $5$ seconds followed by a self-initiated and self-paced reaching movement. Trials with a resting period under 5 seconds were excluded from the evaluation and an error symbol was presented on a monitor for a duration of $200$ \si{\milli\second}. The error symbol consisted of a fixation cross that turned from a green to a red background colour. During the hole experiment, a fixation cross with a green background was continuously shown on the monitor. After each set, the subjects were asked to relax for at least $5$ minutes to avoid any fatigue. The whole experiment was designed and controlled by using the Presentation software [Neurobehavioral Systems, Inc., Albany, USA]. The experimental setup is illustrated in Figure \ref{Figure1}. 

\subsection*{Data Acquisition}

EEG data was recorded using a LiveAmp64 amplifier and an actiCap montage with $64$ active electrodes [Brain Products GmbH, Munich, Germany]. The electrodes were located according to the extended $10$-$20$ system with FCz as reference electrode. All impedances were kept below a threshold of $5$ \si{\kilo\ohm} and were controlled after each measurement set. The data was acquired at a sampling rate of $500$ \si{\hertz} and prefiltered by the measurement device to a bandwidth of $0.1-131$ \si{\hertz}. To avoid possible artifacts during the recording, the subjects were asked to avoid head and eye movements as far as possible. \\
EMG signals were recorded bipolar (Ag/AgCl electrodes) by a WavePlus wireless system and picoEMG sensors by Cometa [Cometa srl., Barregio, Italy]. The EMG was sampled at $2000$ \si{\hertz} and reduced to a bandwidth of $10$-$500$ \si{\hertz} by filters of the measurement device. The signals were recorded from $8$ muscles for the right and left side of the body which were: M. biceps brachii medial, M. triceps brachii medial M. triceps brachii lateral, M. deltoideus lateral, M. deltoideus anterior, M. deltoideus posterior, M. trapezius pars descendens (upper trapezius) and M. flexor carpi radialis. The skin was prepared with alcohol and electrodes were placed according to anatomical landmarks~\cite{Barbero:2012atlas}.  

To mark the physical movement onsets, an infrared motion tracking system [Qualisys AB, Gothenburg, Sweden] was used in addition to the mechanical hand-switches. In total, $4$ motion tracking cameras (Oqus $300+$) were placed in the shielded cabin to record motion data. To track the motions, $3$ reflecting markers were placed on the back of the hand, the elbow (next to the lateral epicondyle) and the deltoideus (muscle belly) on each side of the body. The motion tracking data was acquired at a sampling rate of $500$ \si{\hertz}.

All events during the experiments, such as pressing/releasing the hand-switches and the button as well as invalid trials (shown error symbols) were tracked by the EEG system. Additionally, the start and stop of the recordings of each measurement system was recorded by the trigger channels of the EEG system to synchronize all data in the offline analysis.

\subsection*{Estimation of Physical Movement Onset}
For estimating the physical (ground truth) movement onset, the position data tracked by the motion capture system were analyzed and processed in an offline evaluation. Since the executed reaching tasks consisted of moving the hand from a resting position towards the button, the data from the reflective marker of the moved hand was used for the estimation. Note that in a later rehabilitation session, movement onsets will be measured by the exoskeleton~\cite{folgheraiter:2012measuring,Tabie:2017labelling}. Since only healthy subjects participated in the study, it was assumed for the evaluations, that the right arm is affected whereby the left arm is not affected. Therefore, for bilateral movements only the position data of the left hand was selected for estimating the ground truth movement onset. \\
In a first processing step, the EEG and motion capture data were synchronized. Afterwards the position data was re-initialized to the resting position by subtracting the mean position data, calculated from the first second (resting period) of each experiment. In the next step, the absolute distance to the resting position was calculated by computing the euclidean distance from the three-dimensional position data. Additionally, the velocity of the hand was calculated for each timepoint by taking the difference between two consecutive samples of the euclidean distance. The velocity was filtered by a lowpass filter with a cutoff frequency of $4$ \si{\hertz} (butterworth, $4$. order) and normalized to the maximum value of the current trial. The distance and velocity were then combined by multiplication in order to provide an exact estimate of the movement onset. This procedure was chosen to calculate the onset, independent from small position fluctuations (producing high velocity values) or slight variations of the resting position between trials.\\
Starting from the movement period towards the resting period, it was searched backwards for a datapoint with a magnitude below a defined threshold. The search started at the time where the mechanical hand-switch was released since this is assumed to be the movement onset plus the mechanical delay of the device. The threshold was set to $0.6$ \si{\milli\metre} and specified with respect to the resolution of the motion tracking system after calibration. The movement onsets were marked in the EEG data.  

\subsection*{Channel Selection and Reduction}
\label{Channel Selection and Reduction} 

In order to provide a proper transferability of the classifier, we custom selected EEG channels by means of the knowledge about the surface distribution of relevant EEG activity with respect to individual EEG channels. Due to the fact that the LRP can be observed in the hemisphere contralateral to the side of the moved limb (right arm), we custom selected channels for the classifications that were located on the left hemisphere, especially in the area of the motor cortex. 
By this approach, we aim for enhancing the transferability of the classifier, that is trained on evoked EEG potentials from bilateral movement planning to predict unilateral movement intentions by selecting EEG channels related to right arm movements. 

\begin{figure*}[h!]
\centering
\includegraphics[width=165mm]{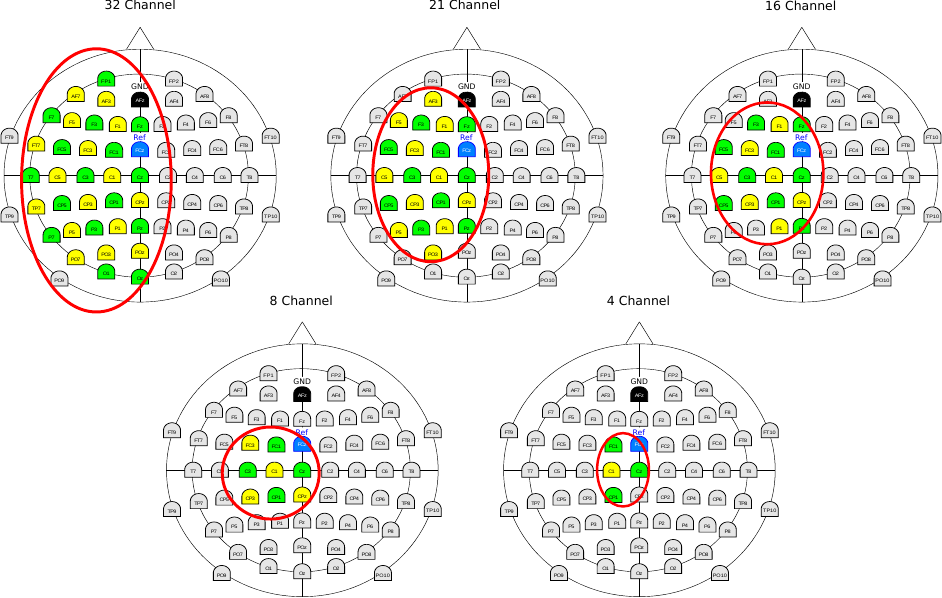}
\caption[Custom selected channels from the left hemisphere.]{Custom selected channels from the left hemisphere. Channels used for the study are marked by a red circle.}  
\label{Figure2}
\end{figure*}

Besides enhancing the performance of the transferred classifier, we also aim to reduce the number of channels to provide an approach that is feasible to be used with persons suffering from stroke. Therefore, we systematically reduced the number of channels used for the classification task in order to reduce the preparation effort in a real rehabilitation session. Due to this we evaluated the use of $32$, $21$, $16$, $8$ and $4$ custom selected channels for the classification task. For all numbers of channels, the selection was made considering the C1 channel as a center of EEG activity related to movement planning, with the other channels located around it. Therefore, by reducing the number of channels the area around this center was further reduced in size. The specified EEG channels for the custom selection are illustrated in Figure~\ref{Figure2}. 

\begin{figure*}[h!]
\centering
\includegraphics[width=165mm]{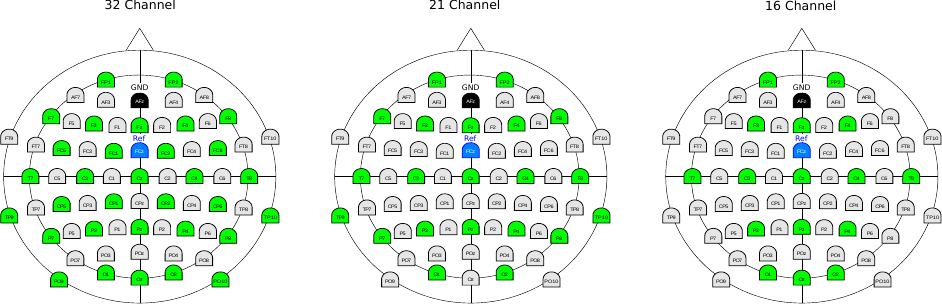}
\caption[Standard channel constellation based on the extended $10$-$20$ system.]{Standard channel constellation based on the extended $10$-$20$ system. Channels used for this study are marked in green.} 
\label{Figure3}
\end{figure*}

In order to evaluate the relevance of custom channel selection, we further compared the custom selection to standard electrode constellations based on the extended $10$-$20$ system. Since such a standard constellation comprises at least 16 EEG channels, we evaluated and compared $32$, $21$ and $16$ channels for the standard constellation as a baseline to our custom channel selection. The standard channel constellations for the different numbers of channels is illustrated in Figure~\ref{Figure3}.

\subsection*{EEG Processing and Classification} 
For the processing and classification of the EEG signals, the signal processing and classification platform pySPACE~\cite{Krell:2013pyspace} was used. A previously developed machine learning pipeline~\cite{kirchner:2014multimodal}, specialized to detect the LRP, was adapted.

\subsubsection*{Preprocessing and Windowing}
The EEG signals were processed window-wise by cutting out overlapping windows with a length of $1$ \si{\second} and a stepsize of $0.05$ \si{\second}. For each trial, a total of $81$ windows, starting from window $[-5.00, -4.00]$ \si{\second} to $[-1.00,  0.00]$ \si{\second} were cut out with respect to the labeled physical movement onset at $0$ \si{\second}. \\
First, a subset of EEG channels corresponding to the evaluated channel selection methods, was included in further processing steps (see Channel Selection and Reduction). Afterwards the data was standardized channel-wise (zero mean, SD of one) and decimated to $20$ \si{\hertz}. Next, a FFT bandpass filter with a passband of $0.1$-$4.0$ \si{\hertz} was applied.

\subsubsection*{Feature Extraction and Classification}
The channel dimension was reduced by applying an xDAWN spatial filter~\cite{Rivet:2009xdawn} with $4$ remaining pseudo-channels, that was designed to enhance event related potentials. Afterwards, the last 4 samples of each window, that correspond to the last $0.2$ \si{\second}, were extracted as time domain features. Therefore, a total of $16$ features were extracted for each window. The features were then normalized by applying a gaussian feature normalization (zero mean, variance one). After feature extraction, an SVM with L1-norm regularization was trained for a binary classification task. The class labels were NoLRP (resting) and LRP (movement intention). The complexity parameter of the SVM was optimized by applying a grid search with $7$ equal spaced values in a range of $10^{-6}$ to $10^{0}$. The class weights of the SVM were set to a ratio of $1:2$ (NoLRP:LRP). The windows $[-1.10, -0.10]$ \si{\second} and $[-1.00, 0.00]$ \si{\second} were used as training instances of the LRP class and the windows $[-3.05, -2.05]$ \si{\second}, $[-3.25, -2.25]$ \si{\second} and $[-3.50, -2.50]$ \si{\second} were selected as training instances of the NoLRP class. After training, the classifier was used to predict all windows of a separated test set. This was done to simulate a real online application scenario, where a classifier is continuously deciding between a resting period and movement intention. The SVM scores were then transformed into a probability by using platts sigmoid function~\cite{Platt:1999probabilistic}. A probability greater than $0.5$ corresponded to a detected movement intention (LRP class), otherwise a resting period (NoLRP class) was detected.

\subsubsection*{Performance Evaluation and Metrics}
Since the class ratios (NoLRP:LRP) are unbalanced for a continous detection of movement intentions (longer resting periods than movement planning) the balanced accuracy (BA) was used as a performance metric. The balanced accuracy calculates the performance with respect to the individual class rates for both classes and is defined as the mean of the true negative rate (TNR) and the true positive rate (TPR). During the evaluation, care was taken that the TNRs and TPRs are not imbalanced in order to avoid a disbalance or bias between the prediction of the NoLRP and LRP classes. \\ 
To emulate an online application scenario, the classifier was evaluated by creating set wise train and test pairs. For each condition, $2$ sets were used for training and the remaining was used as a test set to evaluate the performance. For each condition a total of $24$ performance results were produced due to $3$ train/test permutations for all $8$ subjects. \\ 
To evaluate the performance results in respect to the characteristics of the LRP, a relabelling technique was applied to the classification outcome in order to generate ground truth labels for performance evaluation. Since the individual planning and execution of a movement, for example depending on the waiting time, is affecting the temporal characteristics of the LRP \cite{Schurger:2018specific}, a variability between single trials must be considered. Since the actual start of the movement planning remains unknown, ground truth labels of the windows were computed based on the classification outcome for each individual trial considering an online application. In the following, the procedure is described in detail. 

First, a change point of classes (class boundaries) was computed, which gives an estimate of a started movement planning phase following the resting period and therefore the starting point of the LRP class in time. This change point was defined as a point between two consecutive windows that is determined inside an interval ranging from window $[-2.00, -1.00]$ \si{\second} to window $[-1.00, 0.00]$ \si{\second}. This is the range where the movement planning is to be expected when continously classifying windows in an online application scenario. The windows at the boundary of the defined interval correspond to windows where the true label is known with high certainty for the NoLRP ($[-2.00, -1.00]$ \si{\second}) and LRP ($[-1.00, 0.00]$ \si{\second}) class. The choice of the class boundaries was also discussed in our previous work in~\cite{kirchner:2014multimodal}. If three consecutive NoLRP windows were counted backwards in time (starting from window $[-1.00, 0.00]$ \si{\second} backwards) within this range, the label change point was detected and the labels of all windows prior to this point were set to the NoLRP class and past this point to the LRP class. In case no change point was found inside this range, all windows within this range were defined as instaces of the LRP class corresponding to a detected long movement planning phase. However, for windows where the true label is known (from the experimental design), the class label remained fixed for each movement trial. Therefore, all windows prior to window $[-2.00, -1.00]$ \si{\second} were always instances of the NoLRP class while window $[-1.00, 0.00]$ \si{\second} was always an instance of the LRP class. 

\begin{figure}
\centering
\includegraphics[width=82.5mm]{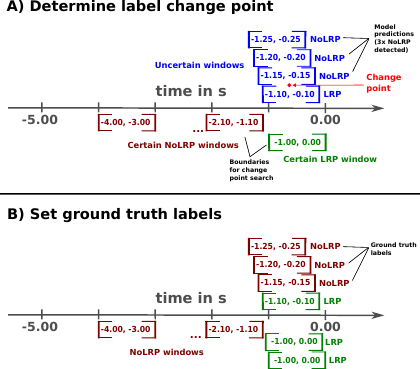}
\caption[Illustration of the relabeling technique]{Illustration of the relabeling technique. In A) the determination of the label change point between two consecutive windows is shown while in B) the ground truth label after applying the method are illustrated.} 
\label{Figure4}
\end{figure}

In conclusion, this technique was used to provide ground truth labels for each sliding window based on the nature of the LRP under predefined constraints where the detection of a movement intention was allowed. To illustrate the procedure, the applied method is shown in Figure \ref{Figure4}. 

\subsection*{Statistical Analysis}
For the main analysis, the classification performances were analyzed by two-way repeated measures ANOVA with \emph{number of channels} and \emph{train-test condition} as within-subjects factors to investigate the effect of transfer depending on the number of channels: \emph{transfer} vs. \emph{no transfer} (see Fig. \ref{Figure5}). Additionally, we performed two-way repeated measures ANOVA with \emph{channel constellation} and \emph{train-test condition} as within-subjects factors to compare both standard constellation and custom channel selection for each \emph{train-test condition} (see Fig. \ref{Figure6}). 

\section*{Results}
\label{sec: results}

\begin{figure*}[h!]
\centering 
\includegraphics[width=120mm]{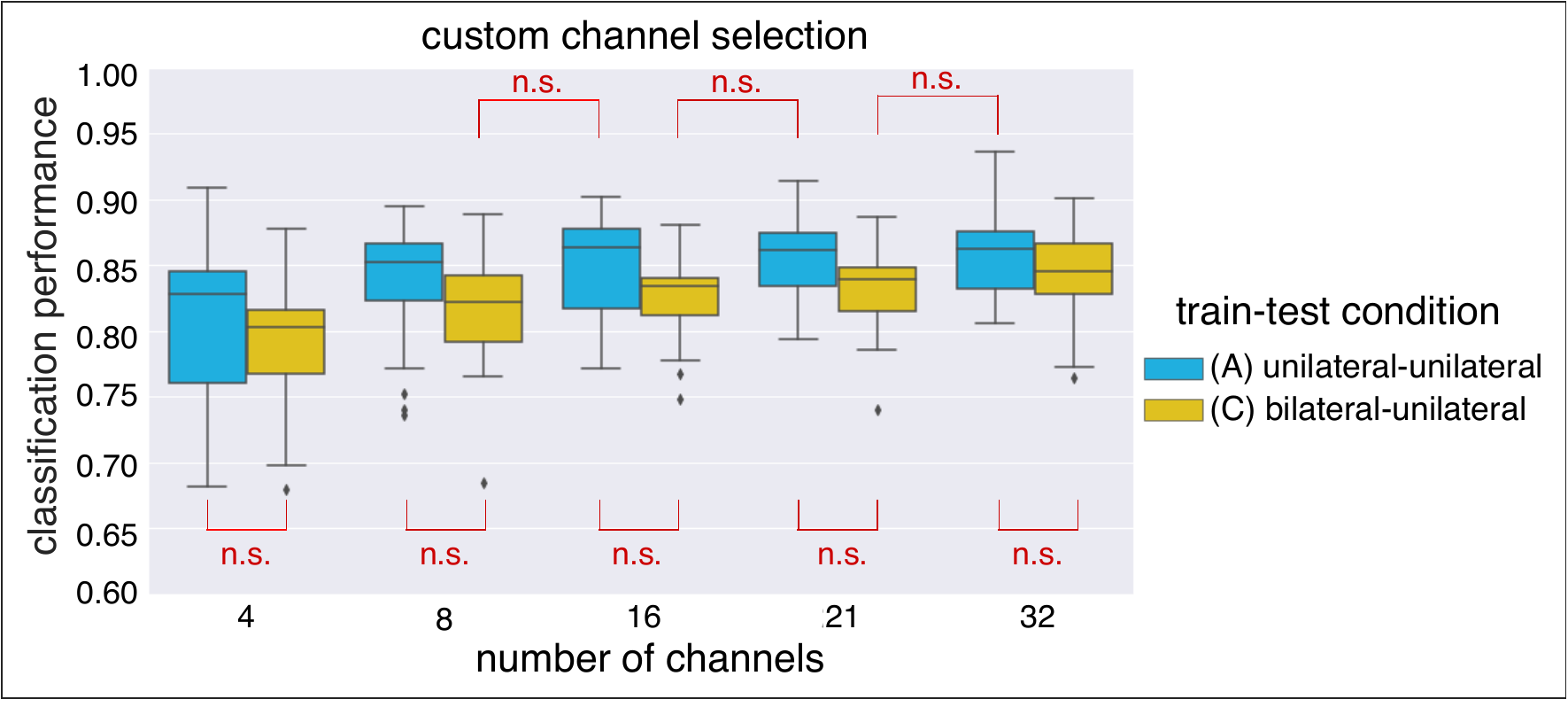}
\caption[CLtransfer]{Transfer effect: classification performance between both \emph{train-test} conditions: (A) no transfer (\emph{unilateral-unilateral}) and (C) transfer (\emph{bilateral-unilateral}). Details for \emph{train-test} conditions, see Table \ref{tab:definition of train-test conditions}. The \emph{n.s.} stands for no significant difference} 
\label{Figure5}
\end{figure*}

Figure~\ref{Figure5} shows the classification performance between both \emph{train-test} conditions: (A) \emph{unilateral- unilateral} (no transfer) and (C) \emph{bilateral-unilateral} (transfer).
We found no significant differences between both \emph{train-test} conditions (A vs. C) for all setups of channel numbers ($32, 21, 16, 8, 4$). 
That means, the classification performance was for the \emph{transfer} case slightly reduced, but did not significantly differ from the \emph{no transfer} case. 

\begin{figure*}[h!]
\centering
\includegraphics[width=170mm]{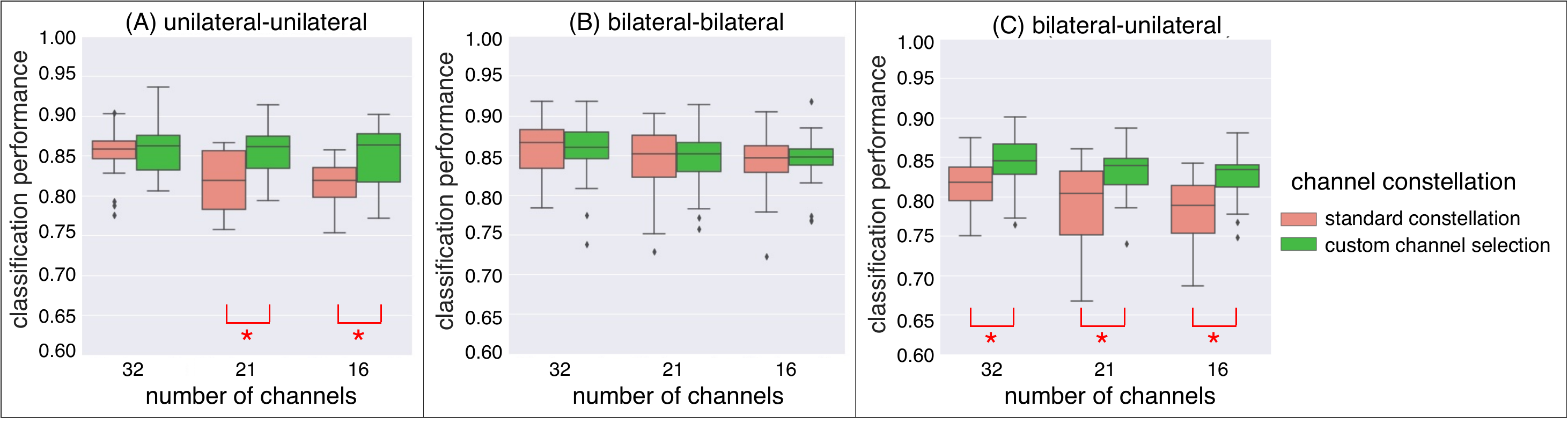}
\caption[elec_distributions]{Effect of electrode distribution: classification performance between standard constellation and custom channel selection for all \emph{train-test} conditions. Details for \emph{train-test} conditions, see Table \ref{tab:definition of train-test conditions}. The $\ast$ stands for significant difference} 
\label{Figure6}
\end{figure*}

Figure~\ref{Figure6} shows the classification performance between both types of channel constellations: standard constellations vs. custom channel selection. The custom channel selection improved the classification performance. This was evident for the case \emph {transfer} [standard constellations vs. custom channel constellations: \emph{n.s.} for all setups of channels, see condition (C) in Fig.~\ref{Figure6}]. The case \emph {no transfer} also benefited from the selection of custom channels when the number of channels was reduced (see condition (A) in Fig.~\ref{Figure6}). However, when EEGs from bilateral movements were used for training and test, we found no differences between both channel constellations, although in the custom channel selection only channels from the left hemisphere were used. Moreover, the classification performance was not affected by channel reduction (see, condition (B) in Fig.~\ref{Figure6}). 

\section*{Discussion}
\label{discussion} 

In this work, we proposed a novel approach to generate labeled EEG data from bilateral movement executions to train a classifier to predict unilateral movement intentions with a high performance. The results show, that unilateral movement intentions can be predicted with a balanced accuracy up to $0.845$ (for $32$ channel, see Fig.~\ref{Figure5}) using the proposed approach for transferring the classifier. This implies that recorded EEG data from a bilateral interaction session, i.e., from a mirror mode rehabilitation session with an exoskeleton, can be used to predict unilateral  movement intentions with a high performance.   

Most interestingly, the results showed that there were no significant differences between classification performances of the condition \emph{unilateral-unilateral} (A) and \emph{bilateral-unilateral} (C) using the custom channel selection although a reduced number of channels was used for the classifications for the transfer condition (C) (see Fig.~\ref{Figure5}). This means, that the proposed transfer approach yields comparable performances to the no-transfer case for a unilateral movement prediction task based on EEG data.

As expected, the results showed that the performance of the classifier systematically decreases with the number of used channels for all conditions. Nevertheless, the results show that the number of channels can be reduced, for example, from $32$ to $21$ channels without a significant performance loss in case of classifier transfer (condition C in Fig.~\ref{Figure5}). This means that a subset of channels covering to some extent relevant brain regions provide sufficiently relevant features for the detection of movement intentions. Even more interesting was, that we did not find significant differences between the transfer condition \emph{bilateral-unilateral} (C) and baseline condition \emph{unilateral-unilateral} (A) even though the number of channels was reduced for example from $32$ to $21$ included channels (see Fig.~\ref{Figure5}). This strongly motivates the applicability of the proposed approach, due to a clear reduction of preparation time when using a reduced number of EEG channels. Nevertheless, the channels must be carefully selected and were specifically chosen in the conducted study depending on the motion task in order to allow the channel reduction without much loss in performance. Therefore, the same number and selection of EEG channels may not be adequate for a different movement task (e.g., hand movements) and an alternative manual or automatic technique can be required when reducing the number of used EEG channels. 

Most importantly, we found that a custom selection of EEG channels outperformed the use of a standard channel constellation for training and transferring the EEG classifier (see Fig.~\ref{Figure6}). Therefore, the custom channel selection outperformed the standard channel constellation, especially for the transfer case. This results indicate, that the custom channel selection allows the possibility to provide a proper transferability of the classifier to predict unilateral movements although only bilateral movements were executed in the training session. Although, preliminary measurements with stroke patients showed that movement intention can be inferred with similar performance compared to healthy subjects, brain activity patterns are different due to the disturbance after stroke. Therefore, the approach has to be further evaluated with EEG data from persons suffering from stroke.

\section*{Conclusion}
\label{conclusion} 
We proposed a novel approach to train an EEG classifier on EEG data recorded during bilateral reaching movements that is afterwards transferred to predict unilateral reaching movements. Classifier transfer was supported by our knowledge-based selection of EEG channels as a data adaptation technique. The approach was evaluated with data from healthy subjects recorded in the conducted study. It was shown that the proposed transfer approach can predict \emph{unilateral} movement intentions with a high performance although the classifier is trained only on EEG data recorded during \emph{bilateral} movements even when using only a small amount of channels. 

Due to the promising results, we are planning further investigations with patients suffering from stroke to evaluate the proposed approach to improve stroke rehabilitation. In future work, we will use our approach in robot-supported rehabilitation using an upper body exoskeleton. In such an application bilateral movements of a hemiplegic patient allows to detect movement onsets from the non-affected arm while the assistive robotic device moves the affected arm synchronously with the unaffected arm (mirror mode). Hence, rehabilitation therapy can take place while EEGs are recorded and labeled to generate training data. After training a classifier on this data, unilateral movement intentions of the affected arm can be detected from it and supported by an exoskeleton. 
However, we expect that in the case of patients suffering from stroke channel selection must be adapted. To this end, an automated approach considering the type and effect of lesion would be preferable. 

We believe that our envisioned novel robot assisted rehabilitation approach can improve future rehabilitation therapy coping with effects of demographic changes such as the rising life expectancy and the accompanying need for more support for the aging population.

\section*{Declarations}


\begin{backmatter}

\section*{Acknowledgements}
We thank Marc Tabie for technical support during the conducted study as well as Kartik Chari for proof reading the manuscript.

\section*{Funding}
This work was funded by the German Federal Ministry of Education and Research (BMBF) within the project EXPECT (Grant number: 01IW20003).

\section*{Abbreviations}
ADL: Assistance in daily living; EMG: Electromyogram; EEG: Electroencephalogram; BCI: Brain computer interface; SSVEP: Steady state visual evoked potentials; TL: Transfer learning; LRP: Lateralized readiness potential; SVM: Support vector machine; BA: Balanced accuracy; TNR: True negative rate; TPR: True positive rate; ErrP: Error-related potential; ERP: Event-related potential

\section*{Availability of data and materials}

The datasets generated and/or analysed during the current study are available in the following Zenodo repository: https://doi.org/10.5281/zenodo.10229480. 

\section*{Ethics approval and consent to participate}
The conducted study was examined and found to be harmless by the University of Bielefeld according to the ethical guidelines of the German Society for Psychology and the Professional Association of German Psychologists.

\section*{Competing interests}
The authors declare that they have no competing interests.

\section*{Consent for publication}
The person on the images used in this manuscript gave a written consent (consent form) for publication.

\section*{Authors' contributions}
N.K. recorded and analyzed the data, conducted the study, and performed the evaluation for classifying the data. E.K. and S.K. designed and supervised the study. S.K. performed the statistical analysis. N.K., E.K., and S.K. have discussed and proposed the classier training approach and equally contributed in writing the manuscript. 

All authors read and approved the final manuscript.

\section*{Authors' information}

N.K. (M.Sc. Biomechatronics) is a researcher with main focus on biosignal analysis (EEG and EMG), machine learning and biosignalprocessing in the field of rehabilitation robotics and robot assisted therapy. 

S.K. is a post-doctoral researcher focusing on brain-computer interfaces, robot learning in human-robot interaction, machine learning, neurophysiological methods (e.g., EEG), and statistics.

E.K. laid the foundation for her interdisciplinary research with a research stay at the Department of Brain and Cognitive Sciences at MIT in Boston/USA. Her research interests are in the areas of human-robot interaction, human-machine interfaces, brain-computer interfaces, embedded brain reading, neurophysiological methods (especially EEG, EMG and other physiological data, and motion data), human behavior analysis, human and artificial agent learning, formal modeling and verification, signal processing, machine learning, embedded AI, embodied AI, and hybrid AI. 
She is a founding member of the "Space2Health" network of the DLR Space Agency. In April 2023, she has been appointed to the "Council for Technological Sovereignty" of the BMBF. 
Furthermore, from 2018 to August 2022 she was a member of Germany’s Platform for Artificial Intelligence in Working Group 6 ‘Health Care, Medical Technology, Care’. In September 2022, she assumed co-leadership of Working Group 7 ‘Learning Robotics Systems’ within this network.


\bibliographystyle{bmc-mathphys} 
\bibliography{bmc_article}      



\end{backmatter}
\end{document}